# Growth of Large-Area and Highly Crystalline $MoS_2$ Thin Layers on Insulating Substrates

Highly crystalline and large-area $MoS_2$ thin layers with good electrical performance can be obtained by the post-annealing of a thermally decomposed ammonium thiomolybdate layer in the presence of sulfur.

*By Keng-Ku Liu[†#], Wenjing Zhang[†#], Yi-Hsien Lee[†], Yu-Chuan Lin[†], Mu-Tung Chang[$],Ching-Yuan Su[%], Chia-Seng Chang[$], Hai Li[@],Yumeng Shi[□], Hua Zhang[@], Chao-Sung Lai[%] and Lain-Jong Li[†&]**

[†] *Institute of Atomic and Molecular Sciences, Academia Sinica, Taipei, 10617, Taiwan*

[$] *Institute of Physics, Academia Sinica, Taipei, 11529, Taiwan*

[%] *Department of Electronic Engineering, Chang Gung University, Tao-Yuan 333, Taiwan*

[@] *School of Materials Science and Engineering, Nanyang Technological University, Singapore.*

[&] *Department of Photonics, National Chiao Tung University, HsinChu 300, Taiwan.*

[□]*Department of Electrical Engineering and Computer Sciences, Massachusetts Institute of Technology, Cambridge, Massachusetts 02139, USA*

[#]These authors contribute equally

To whom correspondence should be addressed: (L. J. Li) lanceli@gate.sinica.edu.tw

**The two-dimensional layer of molybdenum disulfide ($MoS_2$) has recently attracted much interest due to its direct-gap property and potential applications in optoelectronics and energy harvesting. However, the synthetic approach to obtain high quality and large-area $MoS_2$ atomic thin layers is still rare. Here we report that the high temperature annealing of a thermally decomposed ammonium thiomolybdate layer in the presence of sulfur can produce large-area $MoS_2$ thin layers with superior electrical performance on insulating substrates. Spectroscopic and microscopic results reveal that the synthesized $MoS_2$ sheets are highly crystalline. The electron mobility of the bottom-gate transistor devices made of the synthesized $MoS_2$ layer is comparable with those of the micromechanically exfoliated thin sheets from $MoS_2$ crystals. This synthetic approach is simple, scalable and applicable to other transition metal dichalcogenides. Meanwhile, the obtained $MoS_2$ films are transferable to arbitrary substrates, providing great opportunities to make layered composites by stacking various atomically thin layers.**

Graphene holds great promises for replacing conventional Si semiconductors in applications such as high frequency devices and biochemical sensors because of its extremely high carrier mobility and sensitivity to environmental charges.[1,2] However, the zero energy-gap of graphene retards its application in logic electronics. Recently, transition metal dichalcogenides have attracted great attention owing to their two-dimensional (2-d) layer structure analogous to graphene. The transistors fabricated with the molybdenum disulfide ($MoS_2$) atomic thin layers exhibit excellent on/off current ratio and high carrier mobility, which make them suitable for next generation transistors.[3,4] Meanwhile, when the dimension of $MoS_2$ is reduced from a bulk form to a 2-d monolayer sheet, its optical properties change due to the transformation of the band gap from indirect to direct one.[5-11] Significant efforts have been devoted to prepare $MoS_2$ thin layers, including scotch tape assisted micromechanical exfoliation,[3-5,12] intercalation assisted exfoliation,[13-20] solution exfoliation,[15,16,21-23] physical vapor deposition,[25,26] hydrothermal synthesis, [27] electrochemical synthesis,[28] sulfurization of molybdenum oxides[29,30] and thermolysis of the precursor containing Mo and S atoms.[23] However, $MoS_2$ tends to form zero-dimensional closed structures (fullerene-like nanoparticles) or one-dimensional nanotube structures during the synthesis.[31] The method to synthesize large-area and high quality $MoS_2$ sheets with good electrical performance is still rare. Thick $MoS_2$ films (> several tens of nm) have been prepared by spin-coating and thermolysis of alkyldiammonium-thiomolybdate or ammonium thiomolybdate in polar organic solvents.[31-33] However, the carbon contaminations from the residual solvent molecules[32] were suspected to cause the sulfur deficit in the composition (stoichiometric ratio $S/Mo < 2$). The produced $MoS_2$ films are often amorphous or low-crystalline structures. Here, we report a two-step thermolysis process, which is able to grow highly crystalline and large-area $MoS_2$ thin sheets on a variety of insulating substrates. Most importantly, the field-effect transistor (FET) devices based on these $MoS_2$ films exhibit high on/off current ratios and excellent carrier mobility values, comparable with those obtained from the micromechanical exfoliated $MoS_2$ thin sheets.

It has been reported that the thermolysis of ammonium thiomolybdates $(NH_4)_2MoS_4$ in an $N_2$ environment resulted in the conversion of $(NH_4)_2MoS_4$ to $MoS_3$ at 120~360°C[34] as shown in the eq. 1, and the conversion of $MoS_3$ to $MoS_2$ (eq. 2) required the annealing at a higher temperature, for example, above 800°C. It was also suggested that the conversion of $(NH_4)_2MoS_4$ to $MoS_2$ was further lowered to ~ 425°C in the presence of $H_2$ gas[34] as described in the eq.3.

$$(NH_4)_2MoS_4 \rightarrow 2NH_3 + H_2S + MoS_3 \qquad \text{(eq.1)}$$

$$MoS_3 \rightarrow MoS_2 + S \qquad \text{(eq.2)}$$

$$(NH_4)_2MoS_4 + H_2 \rightarrow 2NH_3 + 2H_2S + MoS_2 \qquad \text{(eq.3)}$$

To obtain high quality of $MoS_2$ thin film, it is rational to increase the thermolysis temperature. Note that the direct annealing of the $(NH_4)_2MoS_4$ film at 1000°C in an inert gas does not produce good quality of $MoS_2$ likely due to that the transformation process of $(NH_4)_2MoS_4$ to $MoS_2$ involves many steps and these steps may be easily affected by the presence of oxygen. Thus, $H_2$ gas is needed to avoid the oxidation. However, we also notice that the $MoS_2$ decomposes in $H_2$ when the temperature is higher than 500°C. Thus, a two-step process is developed in this study.

Figure 1a schematically illustrates the two-step process for the synthesis of $MoS_2$ thin layers. High purity of $(NH_4)_2MoS_4$ (Alfa Aesar, purity of 99.99%; 0.25g) was added to 20 ml of dimethylformamide (DMF) to form a 1.25 wt% solution. The solution was sonicated for 20 min before use. An insulating substrate such as sapphire or 300 nm $SiO_2$ on Si ($SiO_2$/Si) was first cleaned with a standard Piranha solution ($H_2SO_4$/ $H_2O_2$ ~ 7/3). After 10 min baking on a hot-plate at 80 °C, the substrate was then immersed into the $(NH_4)_2MoS_4$ solution, followed by slow pulling (0.5 mm/s) to form a thin $(NH_4)_2MoS_4$ film. The substrate was then baked on a hot plate at 120 °C for 30 min.

The annealing processes were performed in a home-made furnace system, where the sample in the quartz tube could be quickly moved between a hot zone (center of the furnace) and a cold zone using a magnet. (Figure S1 in supporting information (SI)). The freshly prepared thin $(NH_4)_2MoS_4$ film was placed in the cold zone of the quartz tube flowing with a gas mixture ($Ar/H_2$ flow rate = 4/1; at 1 Torr). When the center of the furnace reached 500 $^oC$, the substrate was moved to the hot zone of the furnace for the first annealing. The chamber was kept at a low pressure (1 Torr) in an $Ar/H_2$ atmosphere (flow rate 4:1) to efficiently remove the residual solvent, $NH_3$ molecules and other by-products dissociated from the precursors. Sixty minutes later, the sample was moved to the cold zone and the gas environment was changed to Ar (or Ar+S) at 500 Torr. Then the sample is moved to the center of the furnace again for the 2nd annealing when the hot zone reached 1000 $^oC$. The film after the first annealing exhibits two characteristic $MoS_2$ Raman peaks (the $E_{2g}$ mode at ~382 and the $A_{1g}$ mode ~405 $cm^{-1}$) as shown in Figure S2 in SI. It has been reported that the $MoS_2$ structure formed at the thermolysis temperature higher than 300$^o$C.[32,34] However, the relatively larger $E_{2g}$ peak width (~10 $cm^{-1}$) and weaker intensity (relative to the substrate Si peak at 520 $cm^{-1}$) suggest that the crystal structure of $MoS_2$ is still not perfect. Note that the $E_{2g}$ peak width for the $MoS_2$ layers obtained by micromechanical exfoliation is around 4~5 $cm^{-1}$ (Figure S3 in SI). The samples after the 1st annealing are subjected to the 1000$^o$C annealing in pure Ar or in the mixture of Ar and sulfur. The sulfur can be used as a protection gas against the oxidation. Note that the sulfur is introduced into the chamber in a powder form and the powders become sulfur vapors at the process temperature. As shown in Figure S2, the intensity of two characteristic Raman peaks significantly increases and the $E_{2g}$ peak width narrows when the film is further subjected to the second annealing, indicating the high temperature annealing at 1000$^oC$ in Ar improves the $MoS_2$ crystal structure. Surprisingly, the addition of sulfur in the second annealing process greatly improved the crystallinity and electrical performance of the $MoS_2$ thin layers. These arguments are also supported by the photoluminescence (PL) and electrical measurements (conductance *vs.* gate voltage) for the $MoS_2$ layers directly formed on $SiO_2$/Si substrates after the first annealing and those subjected to the second annealing in pure Ar or in the environment with sulfur (Figure S2). It is also noted that the $MoS_2$ layer formed with only the first 500$^oC$ annealing does not show any gate dependence (Figure S2).

The obtained $MoS_2$ films are uniform and continuous based on the optical micrograph shown in Figure 1a. In general, the more diluted precursor solution and faster dip-coating result in a thinner $MoS_2$ layer. Here, we aim at large-area and good quality film for electronic applications. The optimized process reported here is able to produce very homogenous $MoS_2$ trilayers across the whole sample; however, $MoS_2$ bilayers are still occasionally found at some locations of the trilayer film. Figure 1b shows the Raman spectra for the trilayer $MoS_2$ sheets grown on sapphire substrates, where the labels (Ar) and (Ar+S) represent the $MoS_2$ sheets separately annealed in pure Ar and in the mixture of Ar and sulfur during the 2nd annealing. The Raman spectra for the occasionally found $MoS_2$ bilayers are also shown. To compare the Raman signatures of synthesized and micromechanically exfoliated $MoS_2$ thin sheets, we have performed the Raman measurements for the micromechanically exfoliated $MoS_2$ thin sheets with various thickness (number of layers), where the Raman spectra are shown in Figure S3 in SI. Figure 1c shows the energies of the two characteristic Raman peaks for the exfoliated $MoS_2$ sheets. The energy difference between two Raman peaks ($\Delta$) can be used to identify the number of $MoS_2$ layers[8,15] and the relation is shown in the bottom curve of figure 1c. The Δ value obtained in the left and right of Figures 1b is ~20 and ~22.3 $cm^{-1}$, respectively corresponding to the bilayer and trilayer $MoS_2$ films based on the results in Figure 1c.

In addition to Raman features, we observe that photoluminescence (PL) is also very sensitive to the quality of the $MoS_2$ layers. The PL spectra in Figure 1d clearly show that the large-area $MoS_2$ trilayer annealed in the presence of sulfur exhibits a stronger PL intensity compared to that annealed in the environment without sulfur. Note that we have examined the $MoS_2$ films obtained at the temperature lower than 1000°C such as 800, 900 and 950 °C during the second annealing. Our spectroscopic measurements suggest that the crystal structures of these films were poorer than that obtained at 1000°C. The quality of $MoS_2$ layer is also determined by the substrate used for $MoS_2$ growth. We prepared the $MoS_2$ trilayers grown on sapphire and $SiO_2$/Si separately, and then transfer both samples on to freshly cleaned $SiO_2$/Si substrates for spectroscopic measurements. The Raman and PL spectra in Figure S4 in SI consistently show that the quality of the $MoS_2$ layers grown on sapphire is superior than that grown on $SiO_2$/Si. The electrical performance (electron mobility and on/off current ratio) shown in Figure S4 also strongly supports this conclusion. It is known that $SiO_2$ is less stable than sapphire at high temperature; thus, it is suspected that the oxygen from $SiO_2$ may react with the $MoS_2$ film, resulting in lower quality of $MoS_2$ layers. In general the PL and Raman peak intensities may be used to indicate the crystalline quality of $MoS_2$ thin layers.

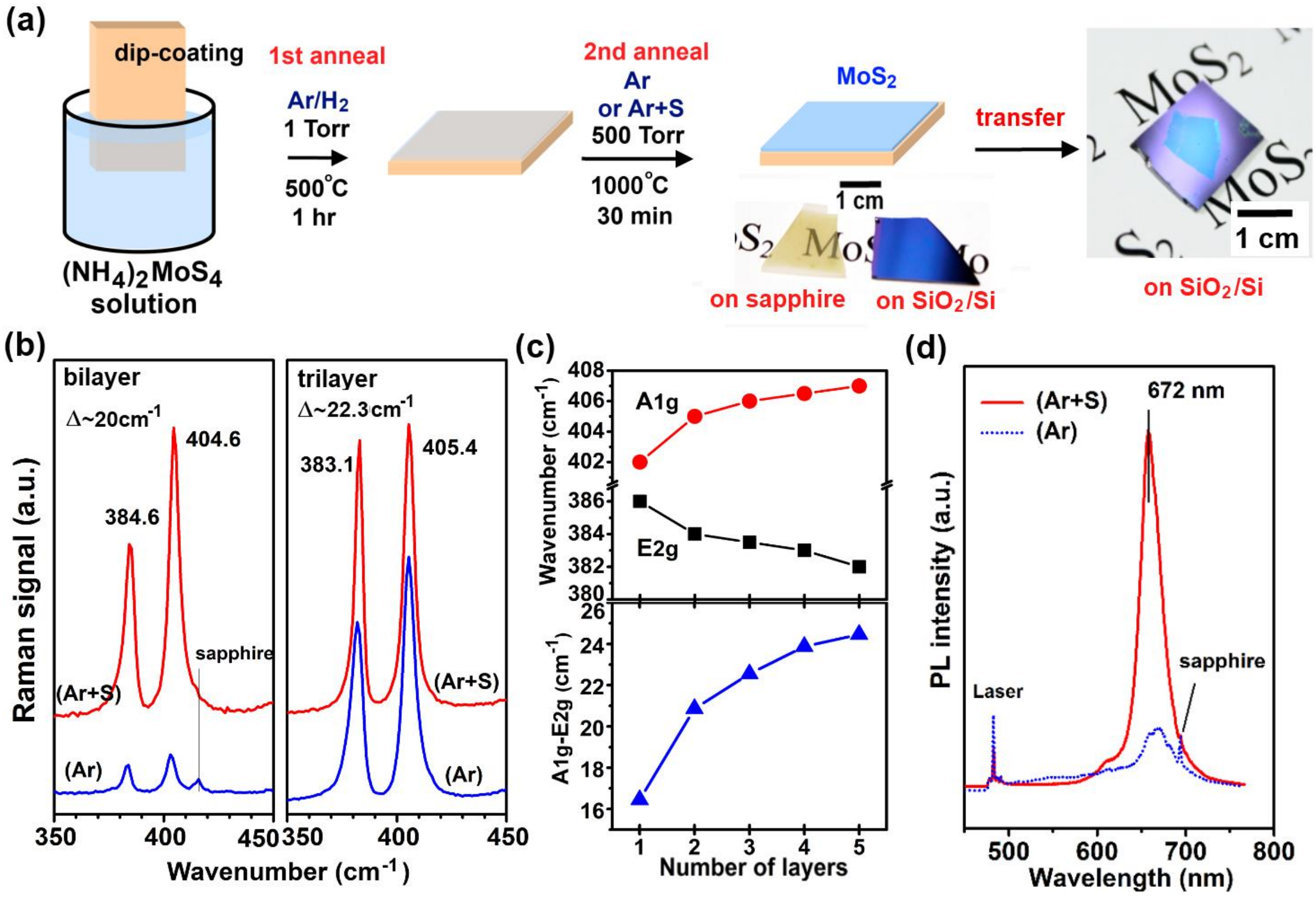

**Figure 1.** (a) Schematic illustration of the two-step thermolysis process for the synthesis of $MoS_2$ thin layers on insulating substrates. The precursor $(NH_4)_2MoS_4$ was dip-coated on $SiO_2$/Si or sapphire substrates followed by the two-step annealing process. The as-grown $MoS_2$ film can be transferred onto other arbitrary substrates. (b) Raman spectra for the bilayer and trilayer $MoS_2$ sheets grown on sapphire substrates (excitation laser: 473 nm), where the labels (Ar) and (Ar+S) represent the $MoS_2$ sheets separately prepared in pure Ar and in the mixture of Ar and sulfur during the second annealing. (c) Energies of the two characteristic Raman peaks for the micromechanically exfoliated $MoS_2$ films with various number of layers. The peak energy difference shown in the bottom graph can be used to identify the number of $MoS_2$ layers. (d) The PL intensity of the trilayer $MoS_2$ thin films prepared in (Ar+S) is stronger than those prepared in pure Ar (Excitation laser 473 nm; Spectra were normalized by Raman scattering peak at around 482nm).

The atomic force microscope (AFM) topographic image in Figure 2a for the obtained trilayer $MoS_2$ film after the second annealing with the presence of sulfur reveals that the film is uniformly flat and with a thickness around 2 nm, where the measured thickness is in agreement with the reported value for exfoliated trilayer $MoS_2$.[3,8,35] The High-resolution tunneling electron microscopy (HRTEM) image in Figure 2b clearly reveals the periodic atom arrangement of the $MoS_2$ film at a selected location. The inset displays the low-magnification TEM image for the folded edge of the $MoS_2$ film (different area in the same sample), where three layers of $MoS_2$ are clearly identified. Note that it is typical that the edge of free-hanging $MoS_2$ sheets folds on a TEM grid after TEM sample preparation process. Thus, the number of layers of the $MoS_2$ film can be resolved under top-view TEM. This method has been used for identifying number of graphene layers.[36] The TEM image in Figure 2c demonstrates that the $MoS_2$ film is highly crystalline. Figure 2d shows the selected area electron diffraction (SEAD) pattern taken with an aperture size (~160 nm) for the sample, as detailed in Figure S5 in SI. Figure S5d also shows the simulated SAED pattern with the $MoS_2$ lattice parameters: hexagonal $P6_3/mmc$ symmetry group, a = b = 3.1 Å, C = 12.8 Å. Note that the lattice constants we use here are obtained from our XRD data in Figure 3a, and these lattice constants are in good agreement with the literature.[37] The distinct hexagonal lattice structure suggests that the film is highly crystalline $MoS_2$. It is noteworthy pointing out that the trilayer $MoS_2$ film synthesized on sapphire with sulfurization is actually polycrystalline and the lateral size of the crystal domain is larger than 160 nm. We have also performed the other comparative measurements. Figures 2e-2f and 2g-2j respectively show the HRTEM and SAED results of the $MoS_2$ layers synthesized at 1000 $^{o}$C with Ar annealing on sapphire and Ar+S annealing on $SiO_2$/Si. It is concluded that the domain size of $MoS_2$ layers grown on sapphire with sulfurization is clearly larger than the size of the films with only Ar annealing (~ tens nm). Moreover, the $MoS_2$ layers grown on a $SiO_2$/Si substrate with sulfurization exhibit inhomogenious crystal grain sizes. Some areas even show non-crystalline structures in HRTEM, as shown in Figures 2i-2j. This observation suggests that the sapphire substrate is better than $SiO_2$.

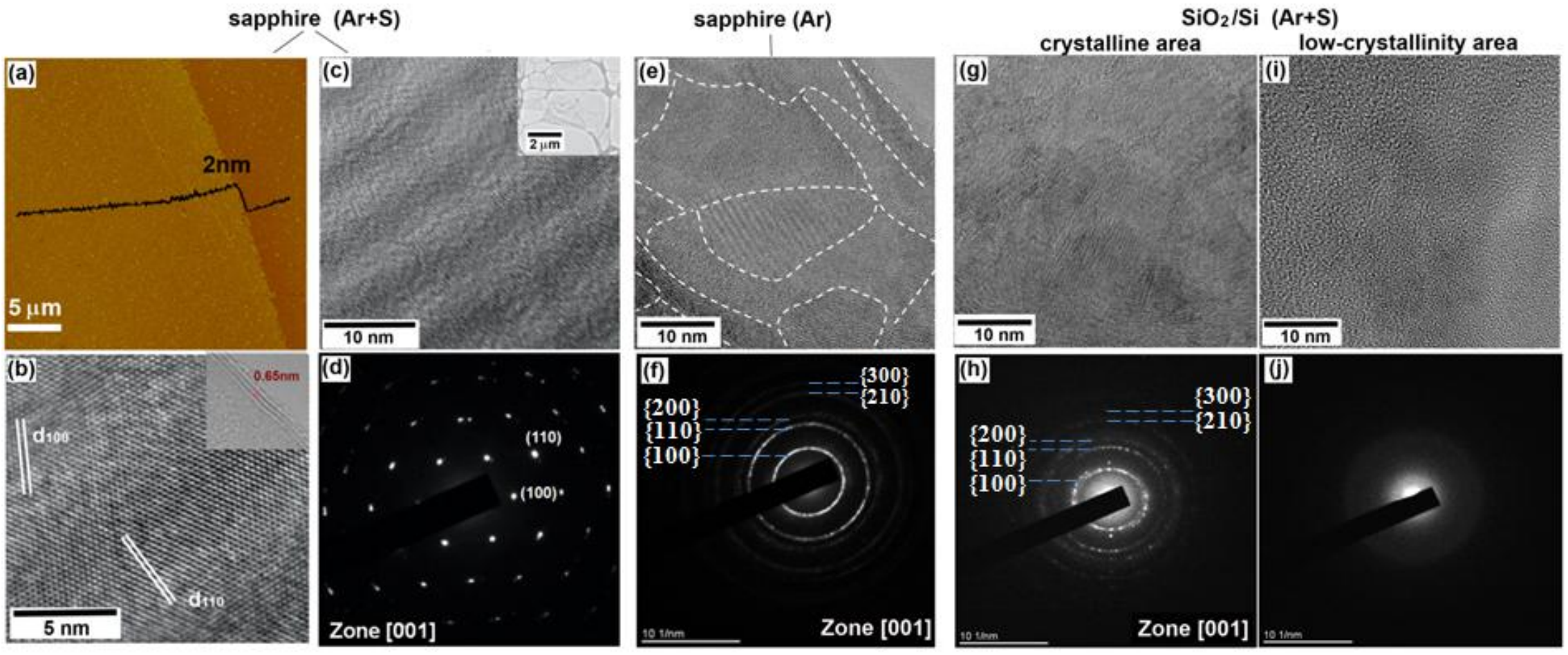

**Figure 2.** (a) AFM image of the $MoS_2$ trilayer grown on a sapphire substrate annealed with the presence of sulfur (Ar+S). (b) High-resolution TEM image for the $MoS_2$ trilayer. The d100 is 0.27 nm, d110 is 0.16 nm. Inset is the TEM image of $MoS_2$ film edge, where three layers of $MoS_2$ are identified. (c) TEM image and (d) the SAED pattern of the $MoS_2$ trilayer sample discussed in (b). (e) TEM image and (f) the SAED pattern for the $MoS_2$ grwon on sapphire with Ar annealing. (g,i) TEM images and (h,j) the SAED patterns for the $MoS_2$ trilayer grown on a $SiO_2$/Si substrate annealed with (Ar+S).

To further reveal the effect of sulfur addition in the second annealing, the glancing incidence angle X-ray diffraction (GIA-XRD) and in-plane X-ray diffraction (in-plane XRD) were used to characterize the obtained $MoS_2$ trilayer. As shown in the schematic illustration in Figure 3a, the detector moves in perpendicular and parallel direction relative to sample surface in GIA-XRD and in-plane XRD, respectively. As a result, the diffraction signals respectively come from the in-plane contribution and the bulk except out-of-plane contribution. Figure 3a shows that the two pronounced peaks at $2\theta = 50.6^o$ and $56.1^o$, assigned as the (105) and (106) reflections respectively, can be clearly identified in GIA-XRD for the sample annealed with sulfur but not for that annealed in pure Ar. These two peaks are related to the diffraction from high-order (00*l*) planes. It is noted that a strong (002) peak is usually observed when the periodicity in c-axis (normal to the $MoS_2$ film plane) is present but the (002) peak can hardly be detected on monolayer or few layer $MoS_2$.[7,13] Both samples show two peaks at $2\theta = 33.1^o$ and $58.9^o$ in the in-plane XRD measurements, assigned to (100) and (110) reflections respectively. Note that all the observed peaks in in-plane and GIA-XRD can be reproduced well with the commonly accepted $MoS_2$ symmetry group: hexagonal $P6_3/mmc$[37].

The addition of sulfur during the second annealing process clearly improved the crystallinity of $MoS_2$ thin layers. The X-ray photoemission spectroscopy (XPS) was used to measure the binding energies of Mo and S in the $MoS_2$ trilayers annealed with and without the presence of sulfur. Figure S6 in SI displays the survey scans for these two films. The detailed binding energy profiles for Mo and S in both cases are similar, where we only show the spectra for the sample annealed with sulfur in Figures 3b and 3c. The Mo3d shows two peaks at 229.3 and 232.5 eV, attributed to the doublet $Mo3d_{5/2}$ and $Mo3d_{3/2}$ respectively. The peaks, corresponding to the $S2p_{1/2}$ and $S2p_{3/2}$ orbital of divalent sulfide ions ($S^{2-}$) are observed at 163.3 and 162 eV. All these results are consistent with the reported values for $MoS_2$ crystal.[14,23] However, the XPS survey spectra in Figure S6 in SI indicate that the oxygen content of the $MoS_2$ annealed in (Ar) is much higher. To further confirm this, TEM based energy dispersive spectroscopy (TEM-EDS) is performed and the results are shown in Table 1. The $MoS_2$ layers annealed in (Ar) exhibit a high oxygen content (22.2 atomic %). The addition of sulfur in the second annealing seems to effectively reduce the oxygen content of $MoS_2$ films to ~ 6.7 atomic %. The oxygen species may come from the moisture absorbed on the $MoS_2$ surfaces and that unavoidably leaked into the chamber from the environments. These oxygen species are suspected to interrupt the crystal structure and degrade the electrical properties of the $MoS_2$. Our experiments suggest that in addition to the increase of grain domain size as evidenced by TEM, the addition of sulfur can also help the removal of excess oxygen species in $MoS_2$ layers.

**Table 1.** Results of TEM-EDS analysis for $MoS_2$ trilayers annealed in (Ar) and in (Ar+S).

| Annealing condition | Atomic percentage Mo | S | O |
|---|---|---|---|
| **(Ar)** | 22.7% | 55.1% | 22.2% |
| **(Ar+S)** | 33.2% | 60.1% | 6.7% |

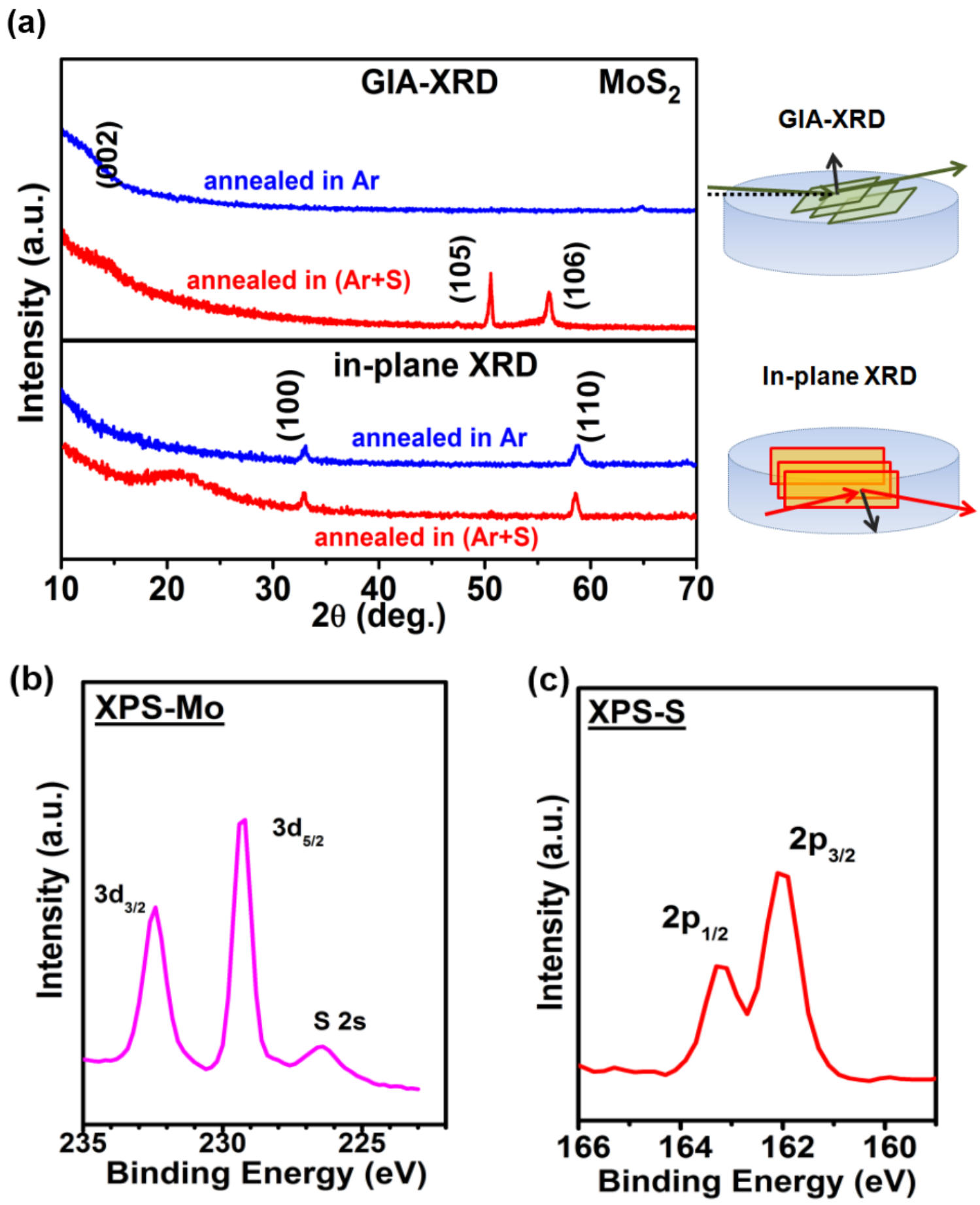

**Figure 3.** (a) The glancing incidence angle X-ray diffraction (GIA-XRD) and in-plane X-ray diffraction (in-plane XRD) patterns of the $MoS_2$ trilayers**.** (b,c) The X-ray photoemission spectroscopy (XPS) measurements for the binding energies of Mo and S in the $MoS_2$ trilayers annealed with sulfur.

To evaluate the electrical performance of the $MoS_2$ thin layers, bottom-gate transistors were fabricated by evaporating Au electrodes directly on top of the $MoS_2$ trilayers on $SiO_2$/Si. Note that the $MoS_2$ films were grown on sapphire substrates and then transferred on to the freshly cleaned $SiO_2$/Si substrates since Figure S4 has demonstrated that the $MoS_2$ films grown on sapphire exhibited much better electron mobility values and on/off current ratios than those grown on $SiO_2$/Si substrates. The transfer process we adopted was similar to the method used for graphene,[38,39] and the details are described in SI. In brief, the detachment of $MoS_2$ from the underlying sapphire substrate was done by partially etching the surface of sapphire with a NaOH solution by control of the etching time and NaOH concentration. Figure S7 in SI displays the AFM images for the as-grown $MoS_2$ films on sapphire substrates and those after transferred onto $SiO_2$/Si substrates. In general, the thickness of the $MoS_2$ films grown on sapphire is quite uniform and the roughness of the large-area $MoS_2$ film is ~0.34 nm and no wrinkles are observed (Figure S6b in SI). After transferred onto $SiO_2$/Si substrates, the film still maintained its initial lateral shape but many wrinkles with a typical height ranging from 5 to 9 nm were formed. Figure 4a shows the top-view optical micrograph of the transistor device. Figure 4b displays the typical output characteristics (drain current $I_d$ *vs.* drain voltage $V_d$) for the $MoS_2$ device. Figure 4c compares the typical transfer curves (conductance *vs.* gate voltage $V_g$) for the devices made from $MoS_2$ trilayers annealed with and without the presence of sulfur. It is observed that the sulfur annealing significantly increases the on/off current ratio from $2.4\times10^3$ to $1.6\times10^5$. Meanwhile, the field-effect electron mobility of the device annealed with sulfur is drastically enhanced from the order of $10^{-2}$ to 4.7 $cm^2$/Vs. Note that the FET shows the typical n-typed behavior, consistent with the other reports.[3,4,35]

The contact metal used in our devices is 80 nm Au with an adhesion layer Ti = 5nm. The Fermi level of Ti (work function: 3.9-4.1eV) is closer to the conduction band edge of $MoS_2$ (4.6-4.9 eV). Therefore, it is anticipated that the FET devices show n-typed behaviors and the threshold gate voltage are negative. The field-effect mobility of electrons was extracted based on the slope $\Delta I_d/\Delta V_g$ fitted to the linear regime of the transfer curves using the equation $\mu = (L/WC_{ox}V_d)(\Delta I_d/\Delta V_g)$, where L and W are the channel length and width and $C_{ox}$ is the gate capacitance.[40] Figure 4d summarizes the statistical distribution of electron mobility values and on/off current ratios for the devices prepared from two types of $MoS_2$ films, where the electrical results agree well with the conclusion drawn from the spectroscopic measurements that the annealing with sulfur is necessary for obtaining high quality $MoS_2$ thin layers. The effective field-effect mobility for the $MoS_2$ device can be up to 6 $cm^2/(V\text{-}s)$ in ambient, comparable with previously reported data (0.1~10 $cm^2/Vs$) for the micromechanically exfoliated $MoS_2$ sheets.[1,3,34,41]

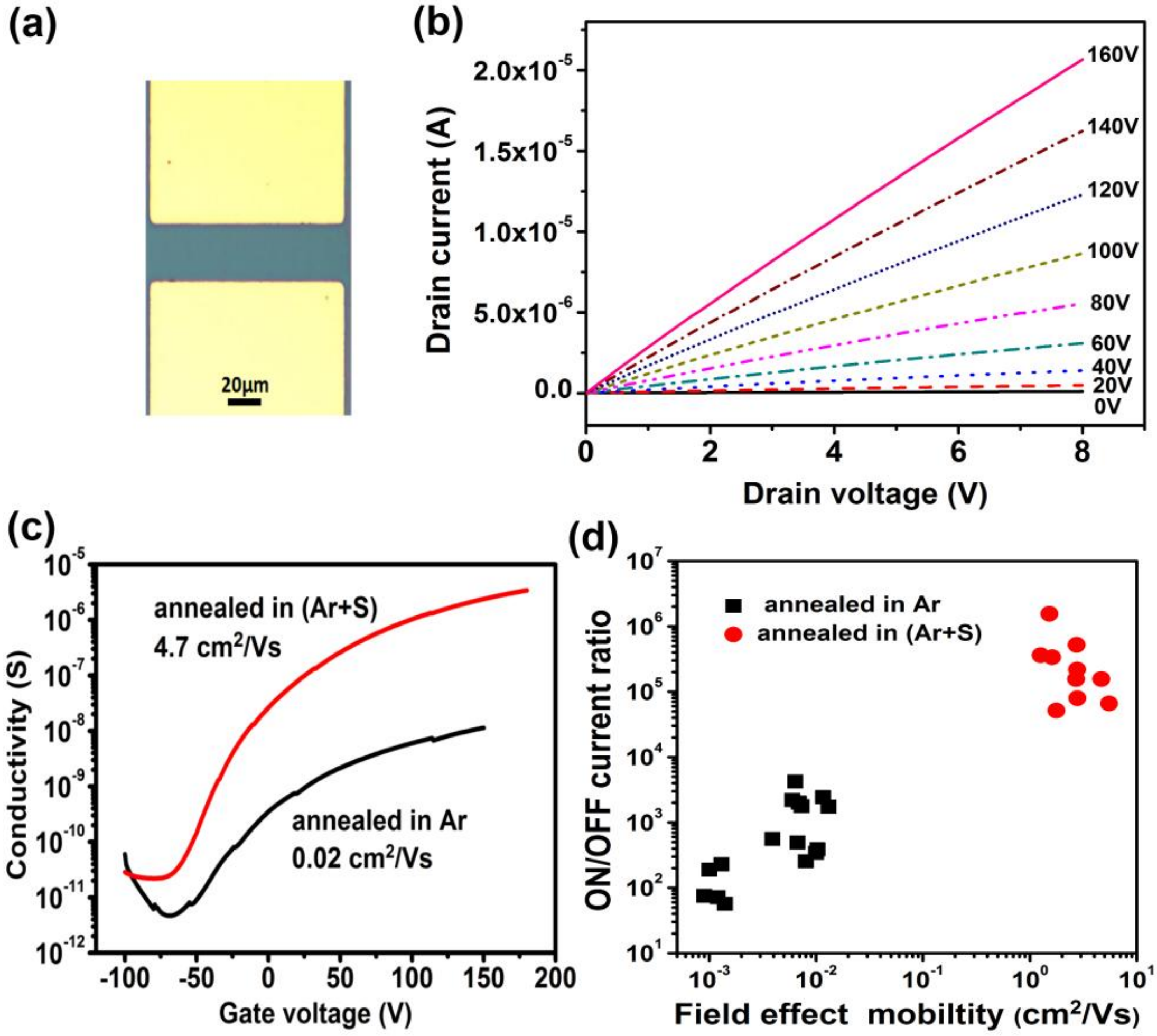

**Figure 4.** (a) Optical micrograph of the top view for the transistor device. (b) Output characteristics of the device made of trilayer $MoS_2$ synthesized on sapphire with (Ar +S). (c) The typical transfer curves (conductance *vs.* gate voltage $V_g$) for the devices fabricated using the $MoS_2$ trilayers annealed with and without sulfur. (d) Statistical comparison of the electron mobility and on/off current ratio for the devices fabricated using the $MoS_2$ trilayers annealed with and without sulfur.

## CONCLUSIONS

In conclusion, we propose a two-step thermolysis process to synthesize large-area and highly crystalline $MoS_2$ thin layers. The addition of sulfur during the high temperature annealing drastically enhances the crystallinity of $MoS_2$, as evidenced by various spectroscopic and microscopic characterizations including Raman, PL, XRD, TEM and SAED. These $MoS_2$ thin layers can be easily transferred onto other arbitrary substrates. The transistor devices fabricated with $MoS_2$ thin layers in a bottom gate geometry exhibit n-type behaviors with the on/off current ratio $\sim 10^5$ and field-effect electron mobility up to 6 $cm^2/Vs$, comparable with the devices prepared by the mechanically exfoliated $MoS_2$. The synthetic approach is simple and scalable.

**Acknowledgements:** This research was supported by Academia Sinica (IAMS and Nano program) and National Science Council Taiwan (NSC-99-2112-M-001-021-MY3 and 99-2738-M-001-001)